\documentclass[floats,floatfix,amsmath,amssymb,prd,twocolumn,superscriptaddress,nofootinbib,preprintnumbers]{revtex4-1}
\pdfoutput=1
\usepackage{graphicx}% Include figure files
\usepackage{dcolumn}% Align table columns on decimal point
\usepackage{bm}% bold math
\usepackage{hyperref}% add hypertext capabilities
\usepackage[mathlines]{lineno}% Enable numbering of text and display math
\newcounter{fnnumber}

\usepackage{color}

\begin{document}

\preprint{APS/EA12173}

\title{Attractors and bifurcation diagrams in complex climate models}

\author{Maura Brunetti}
\email{maura.brunetti@unige.ch}
\author{Charline Ragon}

\affiliation{Group of Applied Physics and Institute for Environmental Sciences, 
University of Geneva, Bd. Carl-Vogt 66, CH-1205 Geneva, Switzerland}

\date{\today}% It is always \today, today,

%MAX 500 words
\begin{abstract}

The climate is a complex non-equilibrium dynamical system that relaxes toward a steady state under the continuous input of solar radiation and dissipative mechanisms. The steady state is not necessarily unique. A useful tool to describe the possible steady states under different forcing is the bifurcation diagram, that reveals the regions of multi-stability, the position of tipping points, and the range of stability of each steady state. However, its construction is highly time consuming in climate models with a dynamical deep ocean, whose 
relaxation time is of the order of thousand years, or other feedback mechanisms that act on even longer time scales, like continental ice or carbon cycle. 
Using a coupled setup of the MIT general circulation model, we test two techniques for the construction of bifurcation diagrams with complementary advantages and reduced execution time. The first is based on the introduction of random fluctuations in the forcing and permits to explore a wide part of phase space. The second reconstructs the stable branches using estimates of the internal variability and of the surface energy imbalance on each attractor, and is more precise in finding the position of tipping points.
\end{abstract}

\keywords{climate system,bifurcations,tipping points,general circulation models}
%Use showkeys class option if keyword
\maketitle

\section{\label{intro} Introduction}

\begin{figure*}[t]
    \centering
    \includegraphics[width=\textwidth]{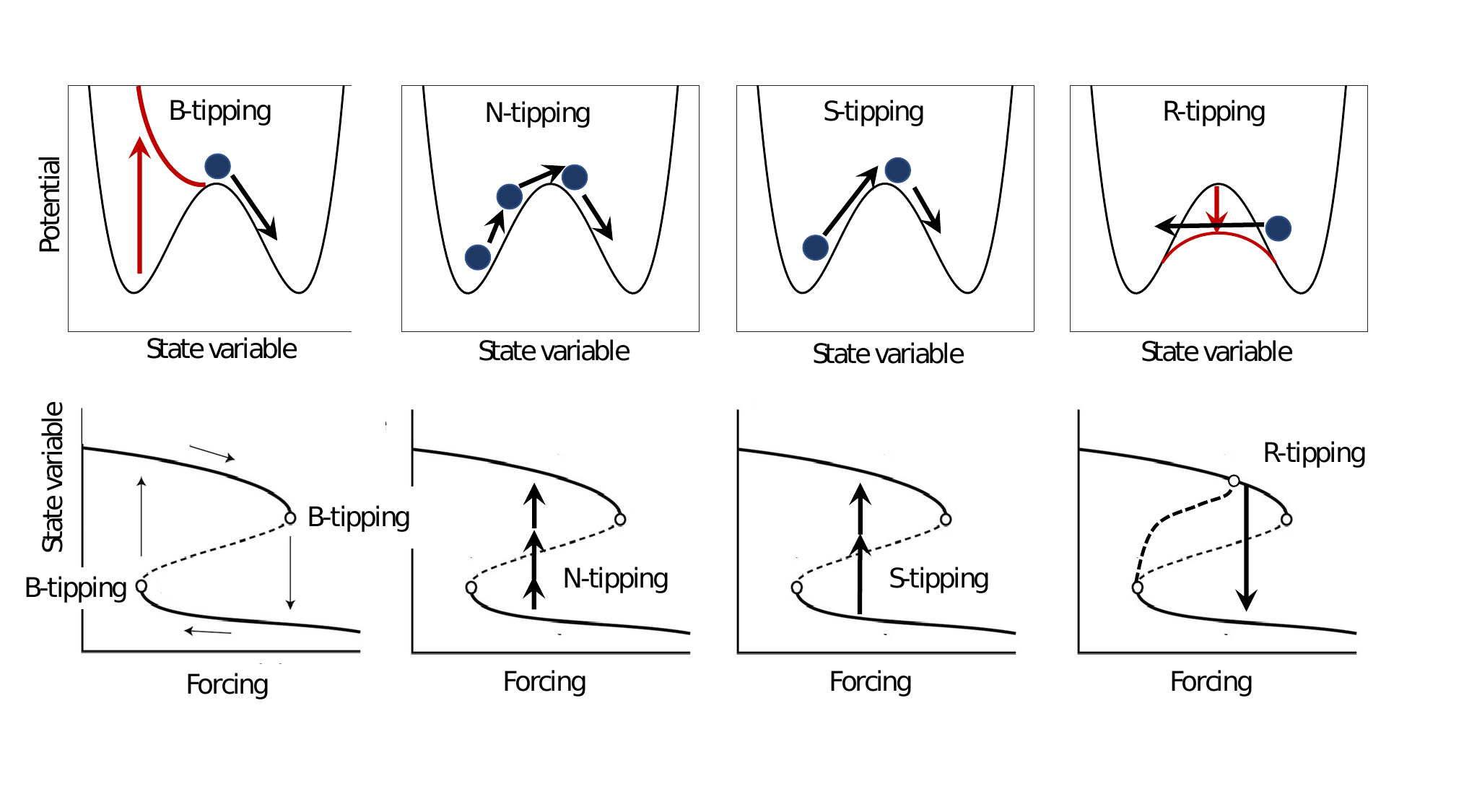}
    %\vspace{-1cm}
    \caption{ \label{fig:Tipping} Tipping mechanisms in terms of potential landscape ({\it first row}) and corresponding bifurcation diagram ({\it second row}). Blue circles: state of the system; red/black arrows: evolution of  landscape/state.}
\end{figure*}

The climate is a dynamical complex system that is fuelled by the incoming solar radiation
and reaches a steady state under the effect of dissipation over a multitude of temporal and spatial
scales~\cite{RevModPhys.92.035002}. Under a given forcing (such as astronomical Milankovitch cycles, or the increasing atmospheric CO$_2$ content due to volcanism or present-day anthropogenic emissions)
the system can be driven out of the steady state.
In such conditions, the system can reach critical thresholds (or tipping points) where its properties abruptly change, often in an irreversible way. 
Such behaviour can be illustrated by the
so-called bifurcation diagram (BD), where a state variable (for example, the mean surface air
temperature) is plotted as a function of the driving force, as schematically shown in Fig.~\ref{fig:Tipping} (second row), together with the corresponding potential curve (first row) for the different types of tipping mechanism~\cite{ashwin2012,vanselow2022}: 1) Bifurcation-induced tipping (B-tipping), that occurs when the deterministic system dynamics reach a bifurcation~\cite{budyko1969,sellers1969,Ghil1976,lucarini2010,ferreira2011,Rose2015,Ferreira2018GRL,Gupta2019,zhurose2022}; 2) Noise-induced tipping (N-tipping), when the internal variability or `climate noise', that occurs in the absence of evolving external forcing and includes processes intrinsic to the system (in the case of climate, to the atmosphere, ocean, land, and cryosphere and their interactions~\cite{deser2012}), increases up to exceed the height of a critical barrier separating two basins of attraction, so that the system can access another dynamical solution~\cite{farrel-abbot-2012,Wordsworth2021,Baum2022}; 3) Shock-induced tipping (S-tipping), related to sudden shocks that induce the passage from one state to another, as happened when a huge asteroid presumably caused the Cretaceous-Tertiary extinction 66 million years ago~\cite{impact1989}, or when volcanic emissions initiated glaciation episodes~\cite{Macdonald2017}; 4) Rate-induced tipping (R-tipping), when the rate of change of the forcing or some internal parameter exceed a critical value~\cite{Ashwin2017,Arnscheidt2020,hoyer2021}, allowing the system to cross, for example, a moving basin boundary (note, however, that R-tipping can be induced by different mechanisms and does not necessarily requires multistability~\cite{vanselow2022}). In climate physics, these mechanisms are extensively studied to illustrate the stability of tipping elements in the present-day climate~\cite{Lenton2008,mckay2022} (two examples among many others are the Arctic sea ice \cite{Eisenman2009,Armour2011,abbot2011a,wagner2015,Hill2016} and the ocean overturning circulation~\cite{stommel1961,Rahmstorf1995,gregory2003,rahmstorf2005,hawkins2011,reviewAMOC2019}), bifurcations at the global scale occurring in the geological past of our planet leading to the snowball state~\cite{roe2010,Voigt2010,voigt2011,Pierrehumbert2011,voigt_abbot_2012,horner2022} or to a state with an ice-free equatorial waterbelt~\cite{abbot2011b,yang2012a,yang2012b,Braun2022}, and also to explore the habitability of exoplanets~\cite{Boschi2013,Checlair2017,Checlair2019}. 

BDs are easily obtained using energy balance models~\cite{budyko1969,sellers1969,Ghil1976}, the simplest in the hierarchy of climate models~\cite{Primer2005}. In intermediate complexity models~\cite{LucariniBodai2017} or low resolution general circulation models~\cite{Rose2015,zhurose2022}, BDs can be still constructed with a reasonable computational cost. As the complexity of the model increases, the amount of CPU time needed to perform series of simulations that explore a huge range of driving force and initial conditions toward steady states becomes prohibitive. Indeed, the standard method requires convergence toward the steady state (or {\it attractor}) to obtain the BD, that means to perform simulations over time scales of the order of several times the relaxation time of the included climatic components~\cite{Primer2005}, which can be $10^3$ years for the deep ocean, $10^4$ years for the carbon equilibration between atmosphere and ocean~\cite{zhurose2022}, and even higher for dynamical ice sheets.  

Using a general circulation model (thus, at the top of the hierarchy in model complexity) in coupled-aquaplanet configuration ({\it i.~e.} a planet entirely covered by the ocean where fully nonlinear interactions are taken into account between atmosphere, ocean and sea ice), Ragon {\it et al.}~\cite{ragon2022} obtained the bifurcation diagram  using the standard method over time scales of thousand years, thus excluding nonlinear feedbacks on longer time scales such as those induced by ice sheets and vegetation. The MIT general circulation model~\cite{marshall_finite-volume_1997,marshall_hydrostatic_1997} (MITgcm) is used here in the same configuration, with coupled atmosphere-ocean-sea ice at 2.8$^\circ$ horizontal resolution, 15 levels in an ocean depth of 3~km, and 5 pressure layers in the atmospheric column. The atmospheric module is based on SPEEDY~\cite{molteni_atmospheric_2003} that provides a rather realistic representation
of the radiative scheme despite the coarse vertical resolution, 
with the advantage of requiring less computer resources than state-of-the-art atmospheric modules. 
On the other hand, the ocean dynamics is as accurate as in state-of-the-art Earth system models and this is crucial to include nonlinear feedbacks on millennial time scales. Using this setup, it is possible to run nearly 200 years in 1 day using 13 cores on clusters like those at the University of Geneva. 
The same MITgcm setup, with an additional land module, has been succesfully applied to investigate the ocean dynamics during the Jurassic~\cite{Brunettiverard2015}, the present-day climate~\cite{Brunetti2018}, and the climatic oscillations in the Early Triassic. 

Here we compare the standard technique to construct the BD with two additional methods which require lower computational costs.  
We will describe such techniques in Section~2, the resulting BDs in Section~3 and we draw our conclusions on pros and cons for each method, and future developments in Section~4. 

%%%%%%%%%%
\section{Construction of bifurcation diagrams}
\label{sec2}

We first describe the standard technique and the corresponding BD. Since, in this case, convergence to the dynamical attractors is required over time scales comparable to the longest feedback process in the setup (in our case, until deep ocean equilibrium is reached), this BD is used for comparisons with those obtained with less computationally expensive techniques that will be described in the following subsections. 

%%%%%%%%
\subsection{Standard technique} 
\label{sec21}

The standard method for BD construction is based on the theory of dynamical systems. The phase space can be divided in basins of attraction, {\it i.~e} the minimal invariant closed sets attracting an open set of initial conditions as time goes to infinity~\cite{Strogatz} for given values of the internal parameters (such as viscosity, diffusion coefficients,  albedo of different surfaces, \ldots) and external forcing. Note that under a constant forcing, the system is ergodic~\cite{tel2020}, {\it i.~e} sufficiently long temporal averages from a single simulation correspond to large ensemble averages. 

The first step is to consider a huge number of initial conditions spanning a large part of phase space. Starting from these initial conditions, the system then evolves in time until the attractors are reached. Statistically steady-state conditions are 
realised within each attractor in the climate system when its global mean annual surface energy balance $F_s$, {\it i.~e.} the sum of sensible, latent, net solar and longwave radiation fluxes at the surface,  becomes nearly zero ($<0.2$~W m$^{-2}$ in absolute value in \cite{brunetti2019}). 
In general, simulation runs over $n\sim 5$-10 times the relaxation time $t_{\text{relax}}$ are needed to guarantee convergence on the attractor~\cite{drotos2017}, where $t_\text{relax}$ depends on the nonlinear feedbacks implemented in the numerical simulation. In an aquaplanet, there are no ice sheets (characterised by a time scale of the order of 10$^5$ yr), and no vegetation (10$^2$-10$^3$~yr). An active carbon cycle between ocean and atmosphere (10$^4$~yr) is also excluded, thus the deep-ocean dynamics is the process with the largest relaxation time of the order of 10$^3$~yr. In our setup, it turns out to be between 500 and 2000 yr, depending on the attractor (see Fig.~2b in \cite{brunetti2019}).  

Five steady climates 
have been found in Brunetti {\it et al.}~\cite{brunetti2019} under the same forcing, represented by the same amount of incoming solar radiation ($S_0=342$~W\,m$^{-2}$) and atmospheric CO$_2$ content (fixed at 326~ppm): {\it snowball} (where ice covers the entire surface), {\it waterbelt} (where an ocean belt survives near the equator), {\it cold state} (with an ice cap extending to 43 degree latitude), {\it warm state} (with an ice cap comparable to the present one, up to 60 degree) and {\it hot state} (a planet without ice). 
Simulations are stopped when the surface energy imbalance $F_{\rm s}$ becomes lower than 0.2~W\,m$^{-2}$ in absolute
value in \cite{brunetti2019}, 
corresponding to an ocean temperature drift 
$dT_\text{o}/dt = F_\text{s}/(c_\text{p} \rho h)$ (Marshall and Plumb 2008, p. 229) lower than
0.05~$^\circ$C per century, with $c_\text{p} = 4000$~J\,K$^{-1}$\,kg$^{-1}$ being the specific
heat capacity, $\rho = 1023$ kg\,m$^{-3}$ the seawater density, and
$h = 3000$~m the ocean depth. 
Indeed, under such conditions,
there is essentially no drift in the annual averages of the climatic
observables.

The second step is to slightly vary the forcing, starting from each attractor, and let the system relax again~\cite{drotos2017}. 
In this way, stable branches of the BD can be constructed. In practise, 
in the aquaplanet simulations, Ragon {\it et al.}~\cite{ragon2022} performed the following: 
the simulations
are initialized from the five attractors, found
in Brunetti {\it et al.}~\cite{brunetti2019} at $S_0 = 342$~W\,m$^{-2}$,  with slightly different
values of the incoming solar radiation until convergence. The CO$_2$ content is kept constant.
This procedure is
repeated until a B-tipping point is reached, that is where a shift of the state variable to a different attractor is observed on a timescale of the order of hundred years. 

Since the attractors are complex dynamical objects living in a high dimensional manifold~\cite{brunetti2019,Falasca2022}, the projection of their invariant (or natural) measure~\cite{[{The natural measure of an attractor describes the probability distribution of the permitted states in phase space, that is how frequently various part of the attractor are visited by the orbit. The natural measure is also said `invariant' under the dynamical system to specify that it does not change under time evolution, as defined in }] Eckmann1985}
on a given state variable, as the surface air temperature $T$, is arbitrary~\cite{FarandaMessori2019,tel2020}. It turns out that different quantities are maximised in each attractor, as found in~\cite{ragon2022}, {\it i.~e.} total precipitation and surface temperature in hot state, intensity of the Lorenz energy cycle in warm state, heat transport in cold state, available potential and kinetic energies in waterbelt, while snowball minimises all the above quantities. Thus, each climatic state would be better represented by a different projection. However, in the present study, the projection is performed, as commonly done in the literature, in terms of the surface air temperature which spans an interval of more than 70~$^\circ$C from the snowball to the hot state, thus differentiating well all the attractors. 

\begin{figure}[t]
    \centering
    \includegraphics[width=0.5\textwidth]{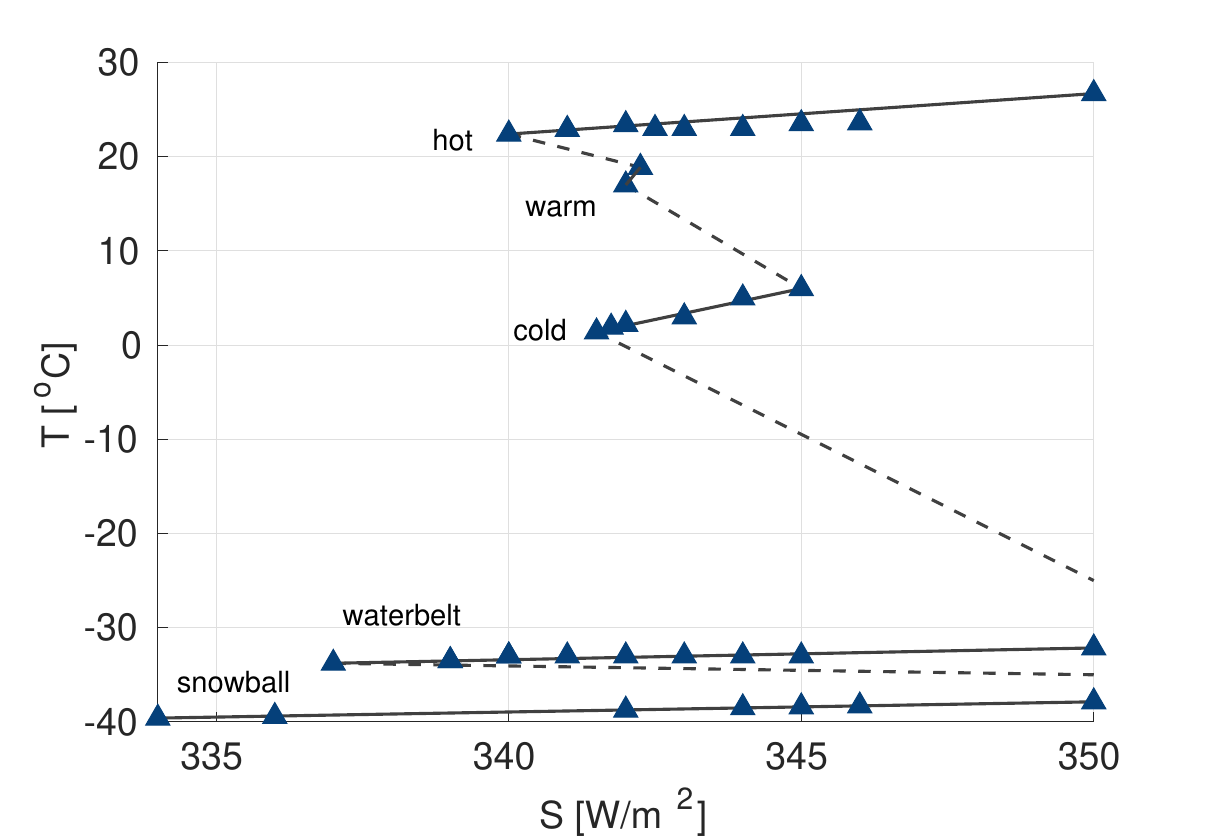}
    \caption{ \label{fig:BDstandard} Bifurcation diagram obtained with the standard method, adapted from \cite{ragon2022}. Solid lines corresponds to stable branches. Dashed lines are a sketch of theoretical unstable branches.}
\end{figure}

The BD obtained in such a way is 
%shown in Figure~1 of Ragon {\it et al}~\cite{ragon2022},
shown in Fig.~\ref{fig:BDstandard}, where it can be seen that the range of stability of the warm state is very small in comparison to the others, while the snowball is stable over all the range of forcing that has been explored, from 334 to 350~W\,m$^{-2}$. The positions of B-tipping are also evident, bracketing either end of the stable ranges for warm and cold states, and the cold end of waterbelt and hot state. Note that the warm state may become unstable when carbon exchanges between atmosphere and ocean are included, as shown in \cite{zhurose2022} using a setup with horizontal resolution of 3.75$^\circ$. 

Such BD has been obtained using a very computationally expensive technique. The total simulation time $t_\text{tot}$ can be estimated as $t_\text{tot} \sim n (N +M)\,t_\text{relax}$,
where $n\sim 10$ to reach convergence, $N \sim 40$ is the number of initial conditions in the first step, $M\sim 40$ is the number of points on the stable branches of the bifurcation diagram (see Fig.~\ref{fig:BDstandard}), giving approximately 
$t_\text{tot} \sim 800\, t_\text{relax}\sim 8\cdot 10^5$~yr. 
Of course, the effective time can be optimized by launching $N$ simulations in parallel with different initial conditions, and by considering separately each stable branch. 
Note, however, that
the term $nMt_\text{relax}$
cannot be reduced further, since the initial conditions at one forcing value are the final state of the previous one. 

%%%%%%
\subsection{Method I: random fluctuations in the forcing}
\label{sec22}

The method is based on the idea  described in \cite{LucariniBodai2020,margazoglou2020} of studying noise-induced transitions (N-tipping) between basins of attraction. In order to explore the entire phase space of the system, and not only the single basin of attraction allowed in nonlinear deterministic dynamics, random fluctuations of the incoming solar radiation are introduced around a given value ($S_0 = 342$~W/m$^2$ in our setup) at discrete times. Even if noise directly affects only a small fraction of degrees of freedom, it propagates to all, since scales are interconnected in the coupled climate model~\cite{LucariniBodai2020}. 

If $\eta$ is a random number taken from a normal distribution with zero mean and standard deviation $\sigma$, a new value of the incoming solar radiation $S$ is prescribed as $S = S_0 (1 + \eta)$ at regular temporal intervals $\Delta t_1$.
We have tested several values of the standard deviation, ranging from $\sigma = 0.01$ to 0.1, and of the temporal interval, from $\Delta t_1 =1$~yr to 100~yr. The smaller $\Delta t_1$, the more uniformly  the phase space is filled, since the system cannot relax toward the attractor, and thus the reconstruction of the position of the attractors in phase space becomes unfeasible. On the contrary, too large values of $\Delta t_1$ are computationally expensive. On the other hand, the standard deviation $\sigma$ regulates the range of forcing values being explored and the portion of phase space than can be accessed, a weaker noise corresponding to rarer transitions between the basins of attraction~\cite{margazoglou2020}. Note that the time step used in our simulations is half an hour  (for the ocean dynamics), much smaller than $\Delta t_1$, thus even if the forcing has Gaussian fluctuations, the resulting time series of the state variable does not correspond to white noise.

In our MITgcm setup, the values $\sigma = 0.025$ and $\Delta t_1  = 10$~yr correctly reproduce the attractors found in the BD with the standard method (Fig.~\ref{fig:BDstandard}) in a reasonable amount of time.
Such values are model- and setup-dependent. For example, in the Planet Simulator (PlaSim), an open-source intermediate complexity climate model with slab ocean~\cite{fraedrich2005}, the relaxation time is governed by upper ocean processes and is of the order of several decades for 5.6$^\circ$ of horizontal resolution and 10 atmospheric levels~\cite{margazoglou2020}. In this case, the reconstruction of the invariant measure is obtained using $\sigma = 0.18$ and $\Delta t_1  = 1$~yr~\cite{margazoglou2020}. We find that a good reconstruction occurs  after a total simulation time of the order of $t_\text{tot} \sim 10^4$~yr. Using different seeds for the random variable and/or changing the initial conditions ($S_0$ or any state variable, possibly using Monte Carlo methods), such total time can be easily partitioned over different set of runs.  

%%%%%
\subsection{Method II: reconstruction of stable branches}
\label{sec23}

The first step coincides with that in the standard technique: the attractors at a given forcing are needed as starting points and are obtained by scanning a large  ensemble of initial conditions~\cite{brunetti2019}. However, the second step differs. We require the change in forcing to be small enough so that the invariant measure on the attractor remains nearly the same~\cite{tel2020}, in particular its mean and standard deviation. 
In other words, while the forcing is changed by $\Delta S$, the corresponding variation on the attractor is small enough that the surface energy imbalance remains nearly zero. This excludes for the presence of R-tipping and guarantees quasi-ergodicity of the system. In practise, 
starting from each attractor found during the first step, the forcing is changed by $+\Delta S$ ($-\Delta S$) each $\Delta t_2=N_2$~yr in order to determine the upper (lower) stable branch in the bifurcation diagram, until a tipping point is reached.

The tipping point is attained when one of the three following criteria is satisfied: 
({\it i}) the standard deviation within the $N_2$ points for each forcing value becomes larger than the internal variability on the attractor (an early warning of critical slowing down of the dynamics~\cite{scheffer2011}); 
({\it ii}) the $N_2$ points turn out to be ordered in time, pointing from the original attractor toward a new one;
({\it iii}) the surface energy imbalance $F_{\rm s}$ becomes much larger than zero (in our setup, larger than 0.5~W m$^{-2}$ in absolute value, which is slightly larger than the threshold 0.2~W m$^{-2}$ used to characterise the convergence to an attractor). When one of these criteria is satisfied, the system is no longer on the initial dynamical attractor and is approaching the unstable branch where a shift toward a new basin of attraction takes place.  

We have tested different values of $\Delta S$ from 0.1 to 0.5~W\,m$^{-2}$ and $N_2$ from 10 to 100. Like in the previous case, the choice is model dependent. We have checked that in our setup good agreement with the bifurcation diagram from the standard technique (that is very accurate albeit highly time consuming) is obtained when $\Delta S \le 0.25$~W\,m$^{-2}$ and $N_2 \geq 20$. With less points, the criterium based on the standard deviation is not applicable (low statistics). With larger $\Delta S$ the position of B-tipping is less accurate and the requirement of remaining on the same original attractor cannot be satisfied, while too small values require large CPU time. 
The total simulation time can be estimated as 
$t_\text{tot} = n N\, t_\text{relax} + M_2\Delta t_2$, where $M_2$ is the number of points in the forcing along all the branches. The second term $M_2 \Delta t_2$ is smaller than the analogous in the standard model, $ nMt_\text{relax}$, since $\Delta t_2 \ll t_\text{relax}$. Note that a small $\Delta S$ corresponds to a large value of $M_2$, thus implying that $\Delta t_2$ can be set to a small value since the system is near the attractor, and Method II can still save time.

%%%%%%%
\section{Results} 
\label{sec3}

We take advantage of the BD obtained with the standard (computationally expensive) technique to test the other two methods described in Sections~\ref{sec22} and \ref{sec23}, and understand under which conditions they can be applied. 

\subsection{Method I}

\begin{figure}[t]
    \centering
    \includegraphics[width=0.5\textwidth]{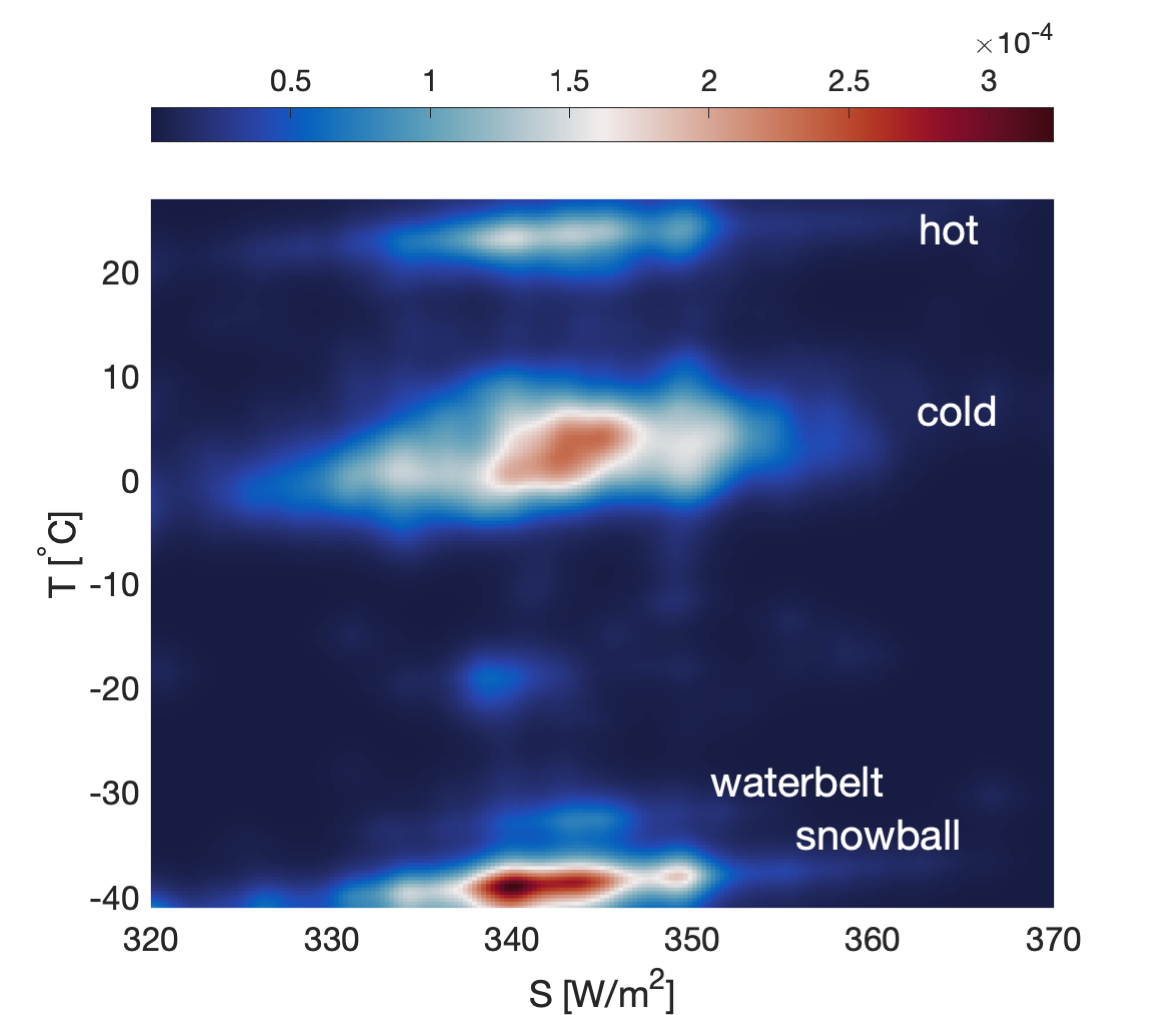}
    \caption{ \label{fig:BDnoise} Normalised 2-dimensional histogram obtained with Method I by adding random fluctuations with standard deviation $\sigma$ = 0.025 to the incoming solar radiation $S_0$ at regular temporal intervals $\Delta t_1 = 10$~yr. The diverging colormap goes from low (blue) to high density of points (red).}
\end{figure}

We construct the bifurcation diagram by plotting the normalised 2-dimensional histogram (projection of the invariant measure) in terms of the global annual surface air temperature and the forcing, as shown in Fig.~\ref{fig:BDnoise} using a standard deviation of the normal distribution equal to $\sigma = 0.025$ and time interval $\Delta t_1 = 10$~yr. Other BDs obtained for different values of $\sigma$ and $\Delta t_1$ are provided in the Supplemental Material (Fig.~S1)~\footnote{See Supplemental Material %at URL 
for additional figures obtained with different parameters in Methods I and II (Figs.~S1, S3),  a description of the transient feature observed around $T \sim -20~^\circ$C in Fig.~\ref{fig:BDnoise} (Fig.~S2), BD in terms of the atmospheric CO$_2$ content at fixed incoming solar radiation (Fig.~S4). \label{SM}}. \setcounter{fnnumber}{\thefootnote}
As can be seen, the overall structure of the phase plane corresponds quite well to that in the standard BD (see Fig.~\ref{fig:BDstandard}). 
The main attractors (hot, cold, waterbelt and snowball) can be easily recognised and correspond to regions of high density of points.  However, the uncertainty in the exact position of the attractors is large, so that, for example, the warm state, with a short stable branch, cannot be distinguished from the hot climate and, in general, it is not possible to precisely infer the edges of the stable branches.

Moreover, a region with increased density different from the ones that correspond to the five attractors appears at $T\sim -20$~$^\circ$C that may correspond to an additional steady state. This can be checked by performing simulations without noise starting from initial conditions in such region of high density, and let the system relax towards a steady state under fixed forcing, {\it i.~e.} until the surface energy imbalance becomes negligible. It turns out that the feature at $T\sim -20$~$^\circ$C is only transient, as shown in the Supplemental Material (Fig.~S2)~\footnotemark[\thefnnumber]: the system remains near such value of temperature for nearly 100~yr with a small surface imbalance but eventually is attracted to the waterbelt.

%%%%%%%%%%%%%%%
\subsection{Method II}

\begin{figure*}[ht!]
    \centering
    \includegraphics[width=0.6\textwidth]{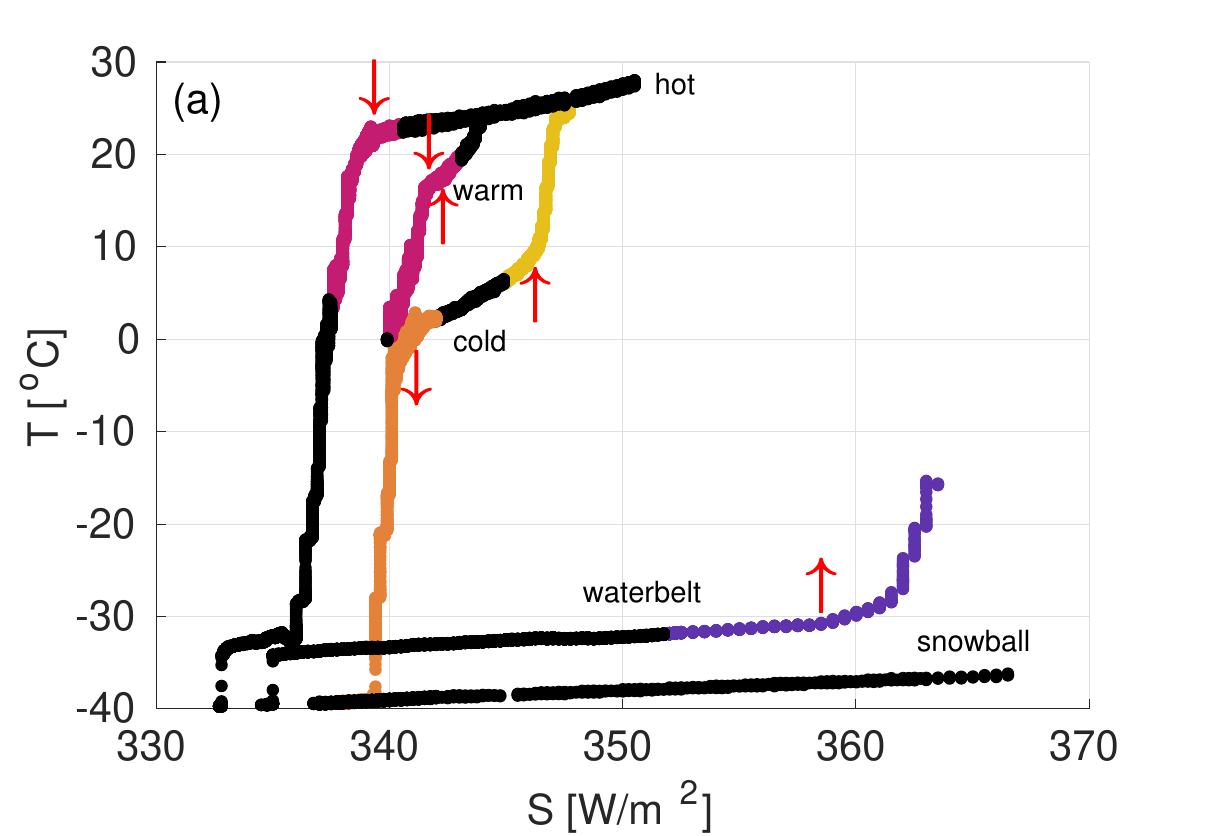}\\
    \includegraphics[width=0.49\textwidth]{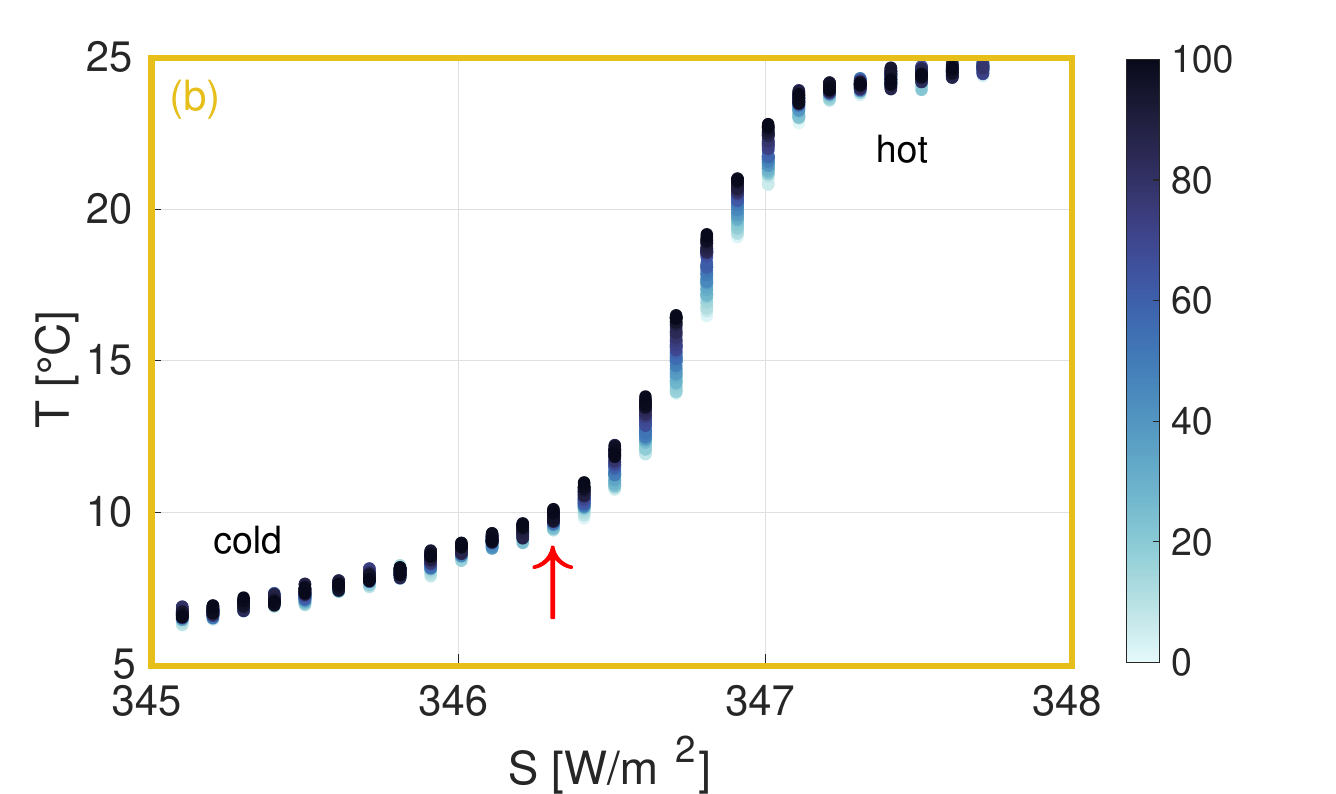}
    \includegraphics[width=0.49\textwidth]{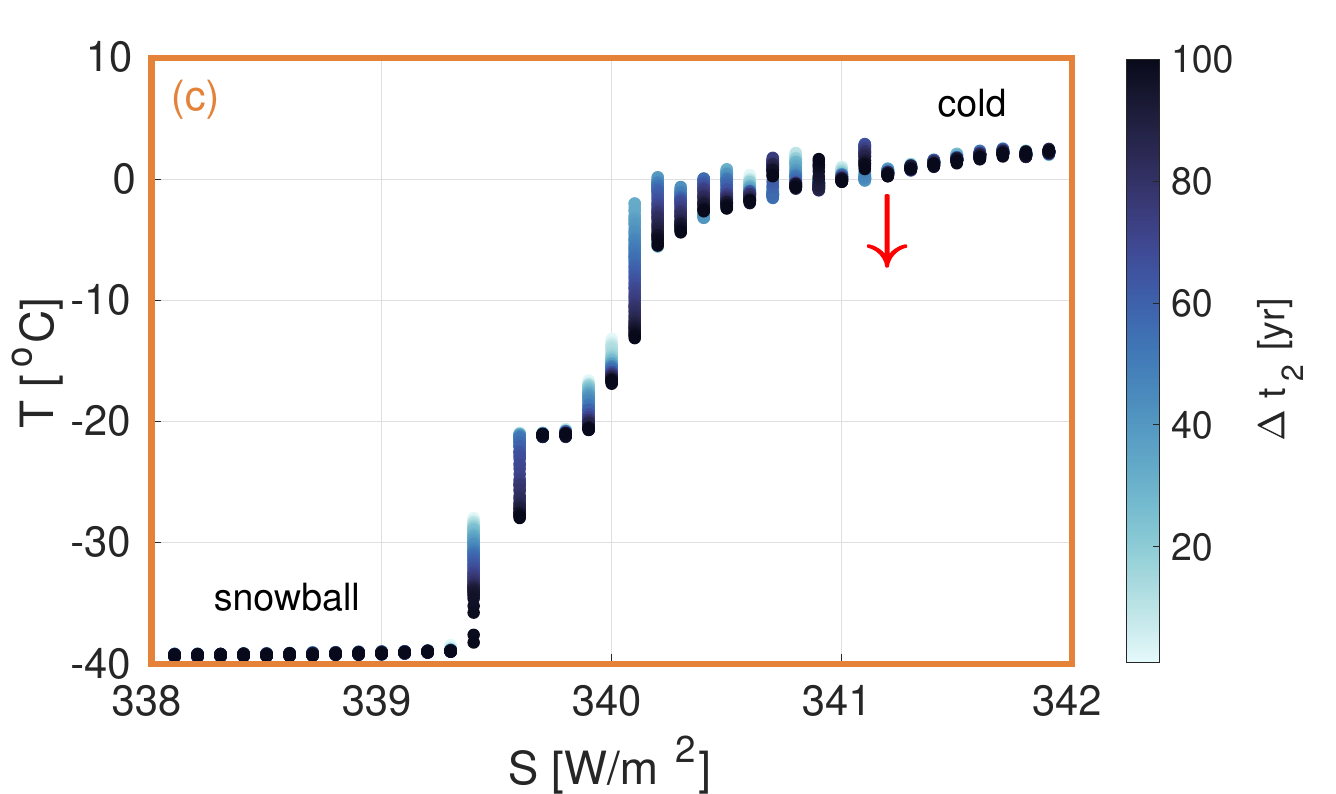}
    \includegraphics[width=0.49\textwidth]{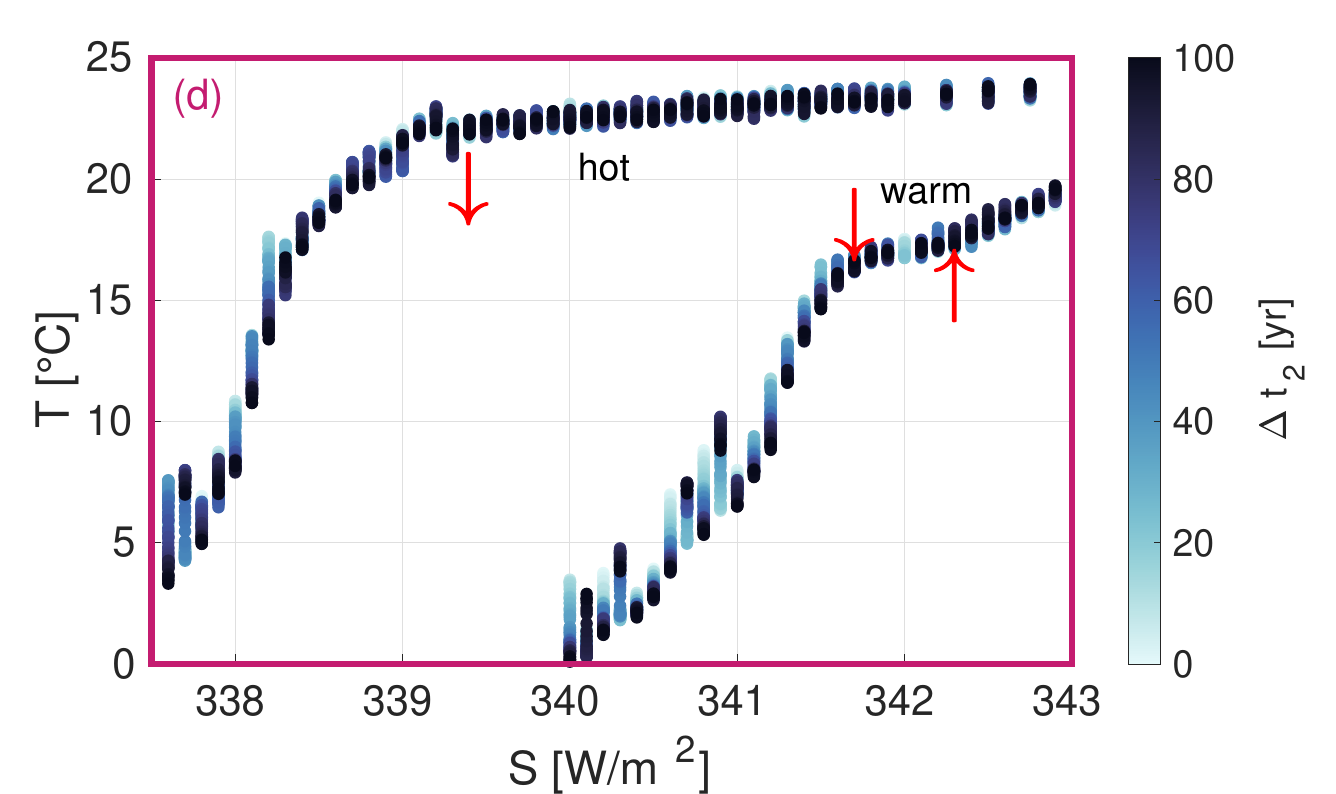}
    \includegraphics[width=0.49\textwidth]{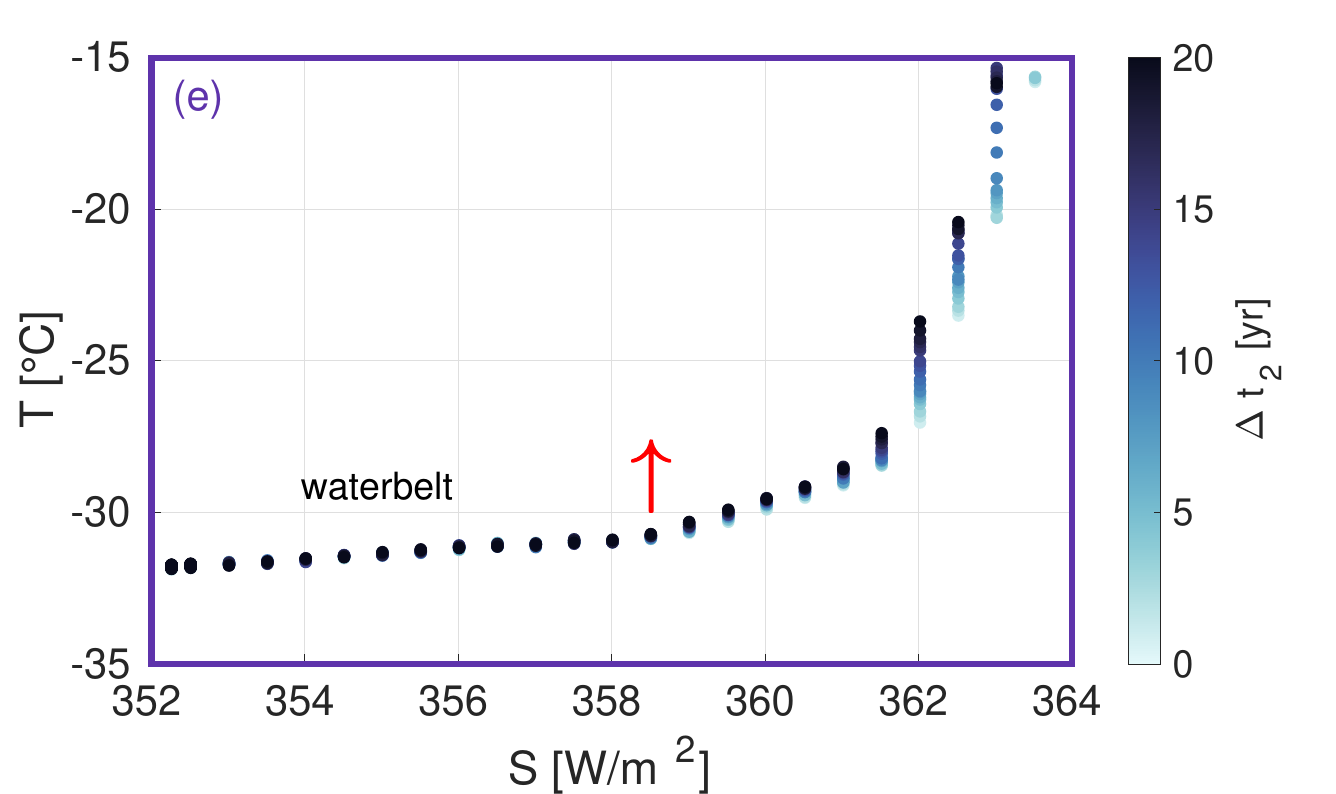}
    \caption{\label{fig:BDbranches} ({\it a}) Bifurcation diagram obtained with Method II; red arrows show the positions of B-tipping obtained with this method and panels {\it b-e} their enlargements, with colors corresponding to the location of tipping from: ({\it b}) cold to hot; ({\it c}) cold to snowball; ({\it d}) warm to cold, warm to hot, hot to colder climates; ({\it e}) waterbelt to hot. Colorbar refers to the time (in years) since the last change in the forcing.}
\end{figure*}

\begin{figure*}[ht!]
    \centering
    \includegraphics[width=0.98\textwidth]{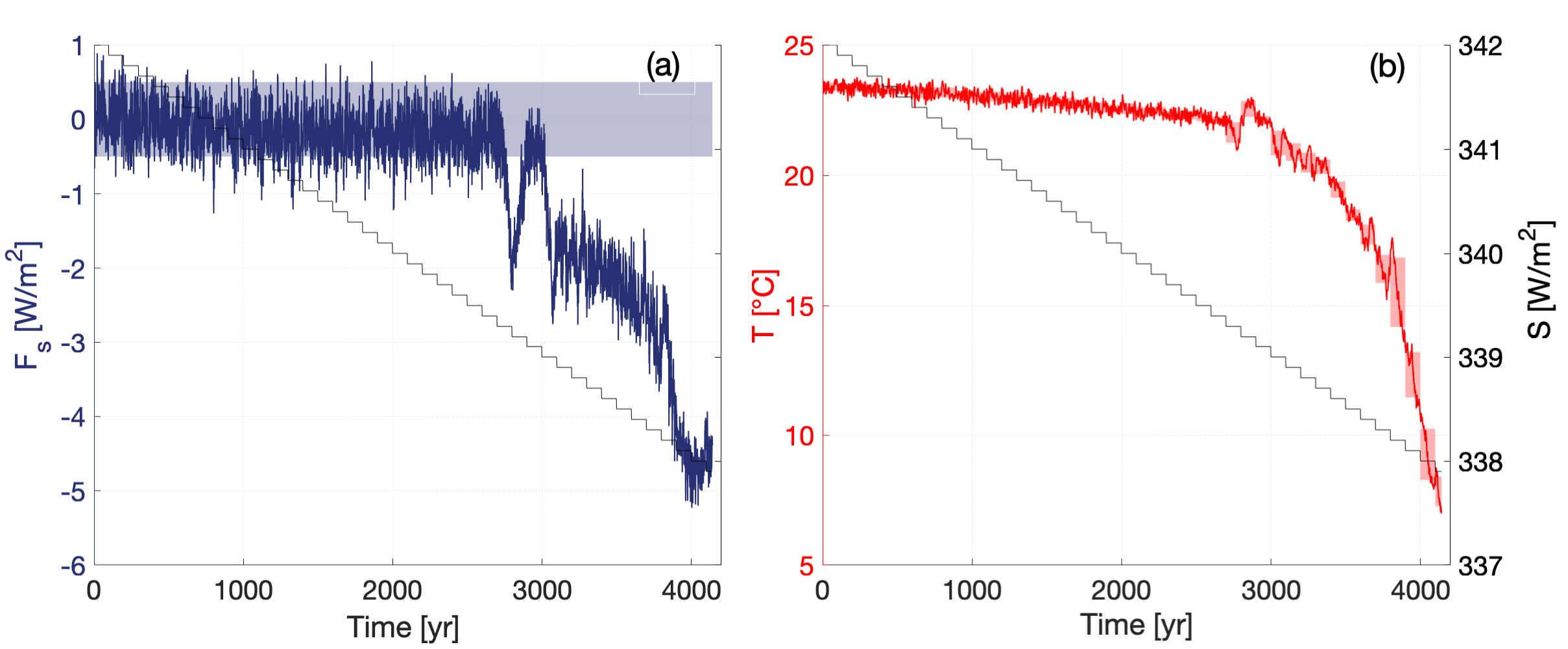}
    \includegraphics[width=0.98\textwidth]{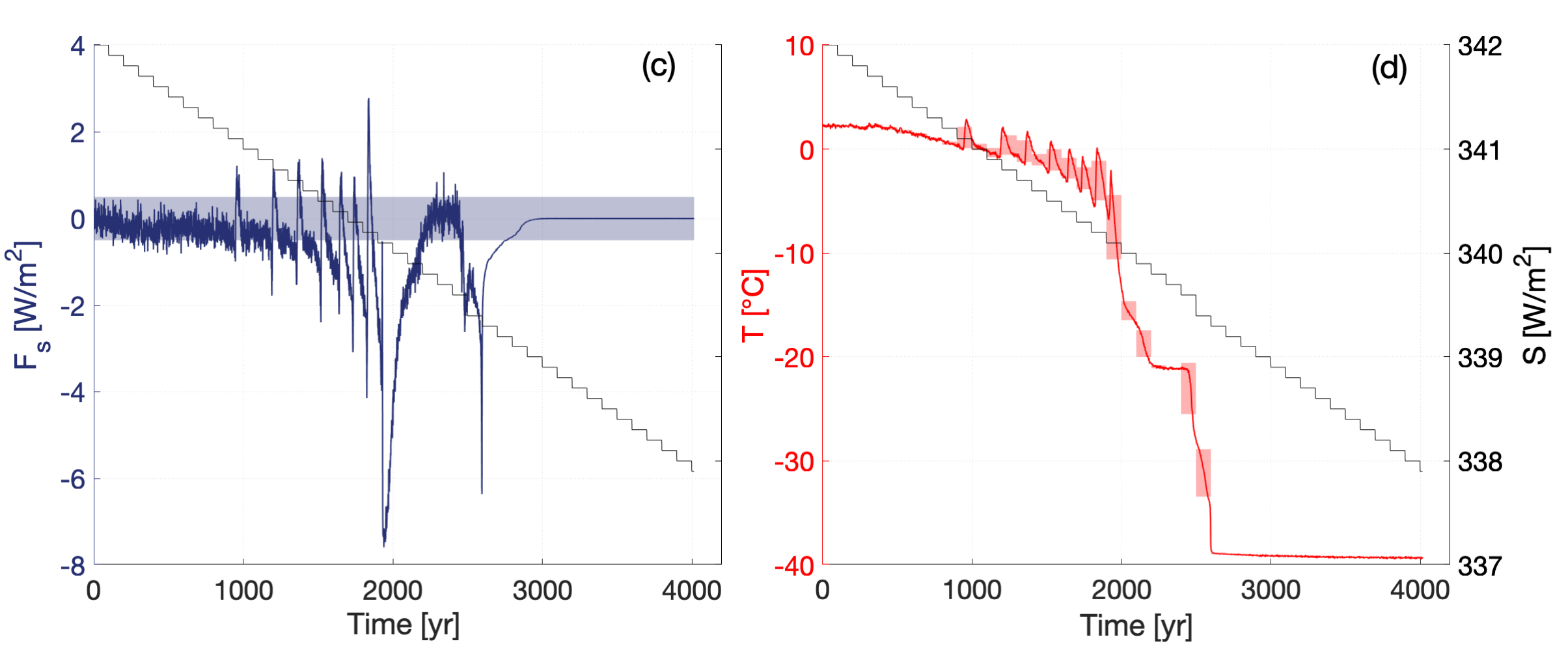}
    \caption{ \label{fig:evolution} Temporal evolution of mean global surface energy imbalance $F_s$ ({\it a,c}) and surface air temperature $T$ ({\it b,d}), with superposed incoming solar radiation (right axes) for ({\it top}) hot to colder climates; ({\it bottom}) cold to snowball. The strip in panels ({\it a,c}) corresponds to an imbalance of $\pm 0.5$~W\,m$^{-2}$. Shadow areas in ({\it b,d}) correspond to the standard deviation in temperature within $N_2 = 100$ points. 
    }
\end{figure*}

Starting from the five attractors found in~\cite{brunetti2019} at $S_0=342$~W/m$^2$, the stable branches are derived by changing the forcing by $\Delta S$ each $N_2$ years. The resulting BD is shown in Fig.~\ref{fig:BDbranches} using $\Delta S = 0.1$~W m$^{-2}$ and $N_2 = 100$ (20 for waterbelt upper branch). As can be seen, the five stable branches can be recovered. If $\Delta S$ is too large, the edges of the stable branches are not correctly reproduced, as shown in the Supplemental Material (Fig.~S3)~\footnotemark[\thefnnumber]. Moreover, by applying the criteria listed in Section~\ref{sec23} to determine the position of B-tipping, we check that they correspond well to those found with the standard technique (see Fig.~\ref{fig:BDbranches}a and its enlargements (panels b, c, d, e, f), where the red arrows correspond to B-tipping and colorbar corresponds to the time ordering of the $N_2$ points). 
In Fig.~\ref{fig:evolution}, two examples of evolution (hot to colder climates and cold to snowball) show when the system abandons the attractor at the point where the standard deviation of the temperature becomes larger than the internal variability on the attractor and/or $|F_{\rm s}|>0.5$~W\,m$^{-2}$ (following the criteria ({\it i}) and ({\it iii}) in Section~\ref{sec23}, respectively).

Another interesting feature in Figs.~\ref{fig:BDbranches}a,c is the transition from the cold state to colder climates. When the system looses stability on the cold branch, it is attracted first by transient structures, like the one at $T= - 20$~$^\circ$C. Then, instead of relaxing to the waterbelt, it is directly attracted to the snowball. This can be understood by remembering that the climate state lives on a high dimensional space that we are arbitrarily projecting to a single state variable (the surface air temperature). In a higher dimensional space, the two climatic trajectories (the transition from cold to snowball and the waterbelt stable branch) would not cross each other. This behavior and the fact that an analogous crossing occurs in the transition from hot to snowball (see Fig.~\ref{fig:BDbranches}a) show that the waterbelt climate is not well described by a single state variable.  

Finally, the dependence of the invariant measure can be investigated, as done here, in terms of the incoming solar radiation $S$ for a fixed atmospheric CO$_2$ content. The opposite can also be considered: different $CO_2$ for fixed $S$~\cite{zhurose2022}. An example is shown in the Supplemental Material (Fig.~S4)~\footnotemark[\thefnnumber] where, however, the MITgcm setup does not include the feedback of the atmospheric carbon with the ocean.

%%%%%%%%%%%%%%%%%%%
\section{Summary and conclusions}
%We have proposed two methods for the construction of BDs
We have tested two methods for the construction of BDs in general circulation models (GCM), where the number and timescale of nonlinear feedback mechanisms make the computational costs of the standard method prohibitive. Such diagrams store crucial information about the nonlinear structure of the climate system, like the regions of multistability, the kind of tipping, the amplitude of climatic oscillations or the intensity of forcing necessary to give rise to a climatic shift. 

In the first method, random fluctuations of the forcing allow to explore the entire phase space of the climate system~\cite{margazoglou2020}. The choice of the parameters for such method, {\it i.~e.} the standard deviation of the Gaussian fluctuations and the temporal interval after which the forcing randomly changes its value, is model dependent. However, once the parameters are fixed, 
such method gives an overall picture of the phase space projected on a given state variable, revealing the number of attractors and their position. The disadvantage is that the attractors with small stable branches may be blurred and the position of B-tippings is not precise. 

The second method reconstructs the stable branches of the attractors~\cite{Rahmstorf1995} by changing the forcing by a small amount that guarantees the system to remain on the same attractor until a B-tipping is reached. The signature of a shift is given by increased variance of the internal variability, non-null surface energy imbalance or an ordered temporal sequence of points toward another climatic state. The advantage of such a method is that the reconstruction of B-tipping is quite accurate with much lower computational costs with respect to the standard method.

The two methods can be used in sequence when no previous information on the attractors is available: a first guess on the position of tipping points and the extension of stable branches can be obtained using Method I. Then, simulations with the standard method or Method II can be performed to fill in the details of the bifurcation diagram, depending on the desired level of precision. 

The proposed methods have been described by projecting the high dimensional space of the attractors to a single state variable (global annual mean of surface air temperature). Of course, they can be applied using other projections, and averages over different temporal and spatial scales. This is particularly useful in order to develop early warning indicators that are based not only on temporal series but also on spatial information~\cite{spatialEW2019,livina2023}. From the BD one can infer the forcing that induces a tipping and perform a detailed analysis in space to identify where the main changes take place, since gridded data in general circulation models permit such kind of spatial studies. The advantage is that spatial early warning indicators do not need a long temporal record to get a meaningful signal, thus they play a crucial role in identifying tipping mechanisms from datasets with irregular or infrequent temporal resolution~\cite{spatialEW2018}.

BDs can also be applied to estimate the climate sensitivity~\cite{Ragone2016,vonderheydt2017,tel2020}, in particular to analyse how this metric depends on the attractor, on the nonlinear feedback mechanisms included in the simulations within a given integration time, on the phase space region explored by the considered initial conditions, and on the perturbation amplitude, all aspects recently discussed in~\cite{bastiaansen2023}.
 
Apart from the fundamental and practical interest of obtaining the BD for our present-day climate, there are many aspects that are worth to be analysed. An open question is, for example, whether the attractors and the BD are model dependent. 
The robustness of the results presented here should be tested against other climatic models and other BD reconstruction methods~\cite{LucariniBodai2020,zhang2021}. For the comparison, analogous numerical setups should be chosen at the level of horizontal/vertical resolutions and included nonlinear feedback mechanisms. Important information on the description of the nonlinear processes and their interplay can be gained from such comparison of BDs produced by different models, that can be used to improve algorithms and to correct biases. 
This is why we suggest including the comparison of BDs in coarse-resolution GCMs with simplified configurations in the Tipping Point Model Intercomparison Project (\href{https://global-tipping-points.org/programme/breakout-workshops/how-to-advance-modelling-of-climate-tipping-points-tipmip-workshop/}{\tt TipMIP}) beside that of Earth System Models in the present-day configuration.
%\footnote{\url{https://www.wcrp-climate.org/slc-events-opportunities/slc-tipping-points-discussion}}.    

%\acknowledgments
\begin{acknowledgments}
We are grateful to  Sacha Medaer and Enzo Samy Ferrao 
for running some of the MITgcm simulations with noise. We thank Alexis Gomel and J\'er\^ome Kasparian for useful discussions.
The computations were performed on the Baobab and Yggdrasil clusters at
University of Geneva. 
The data that support the findings of this study were generated by the MIT general circulation model that is openly available on GitHub  (\url{http://mitgcm.org/}, \url{https://github.com/MITgcm/MITgcm}, version c67f).
We acknowledge the financial support from the Swiss National
Science Foundation (Sinergia Project CRSII5\_180253).
\end{acknowledgments}

%% ------------------------------------------------------------------------ %%
%% References and Citations
%%%%%%%%%%%%%%%%%%%%%%%%%%%%%%%%%%%%%%%%%%%%%%%
% \bibliography{<name of your .bib file>} don't specify the file extension
%
\bibstyle{natbib}
\bibliography{attractors}

\end{document}

%\title{Supporting Information for "Attractors and bifurcation diagrams in complex climate models"}
\title{Supplemental Material for \\ `Attractors and bifurcation diagrams in complex climate models'}

\author{Maura Brunetti}
\email{maura.brunetti@unige.ch}
\author{Charline Ragon}

\affiliation{Group of Applied Physics and Institute for Environmental Sciences, 
University of Geneva, Bd. Carl-Vogt 66, CH-1205 Geneva, Switzerland}

%\date{}

%\preprint{}

\maketitle

\section*{Content of this document}

\begin{figure*}[h!]
\centering
\noindent\includegraphics[width=0.48\textwidth]{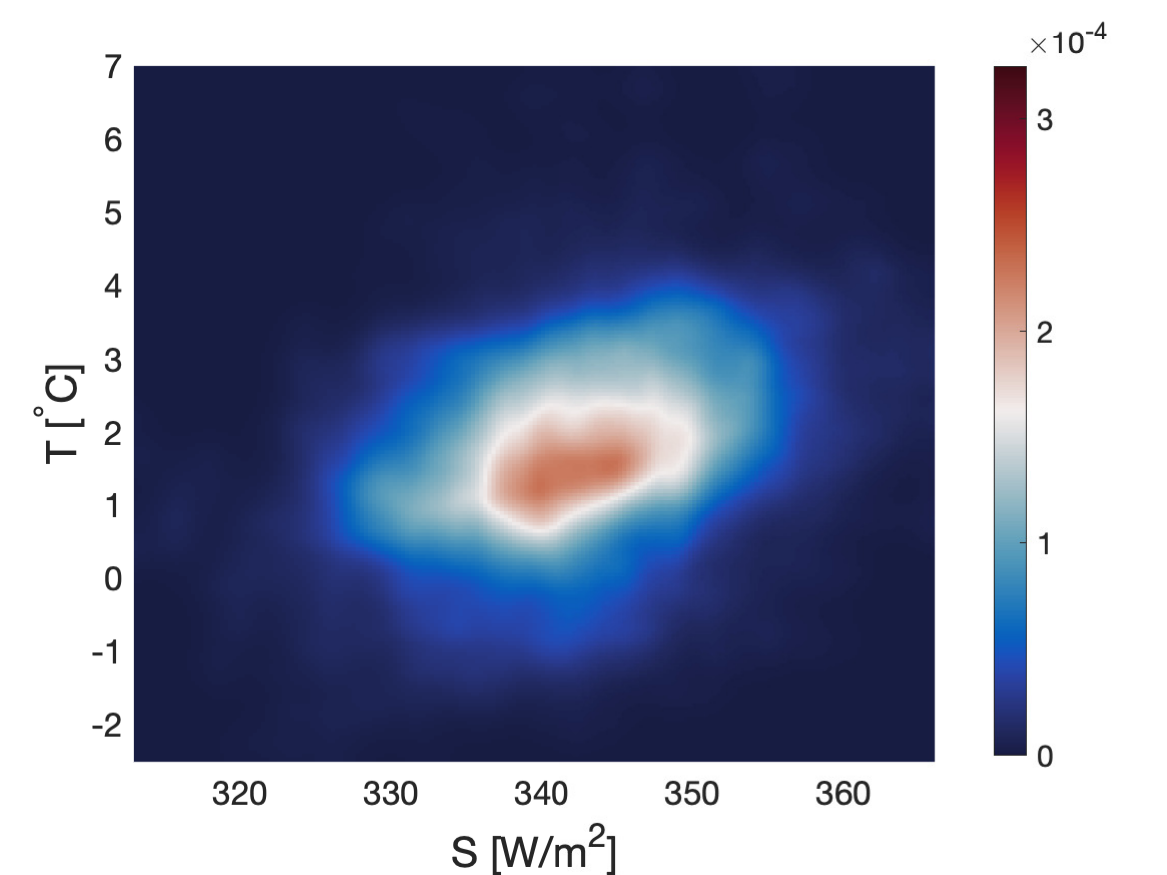}
\noindent\includegraphics[width=0.48\textwidth]{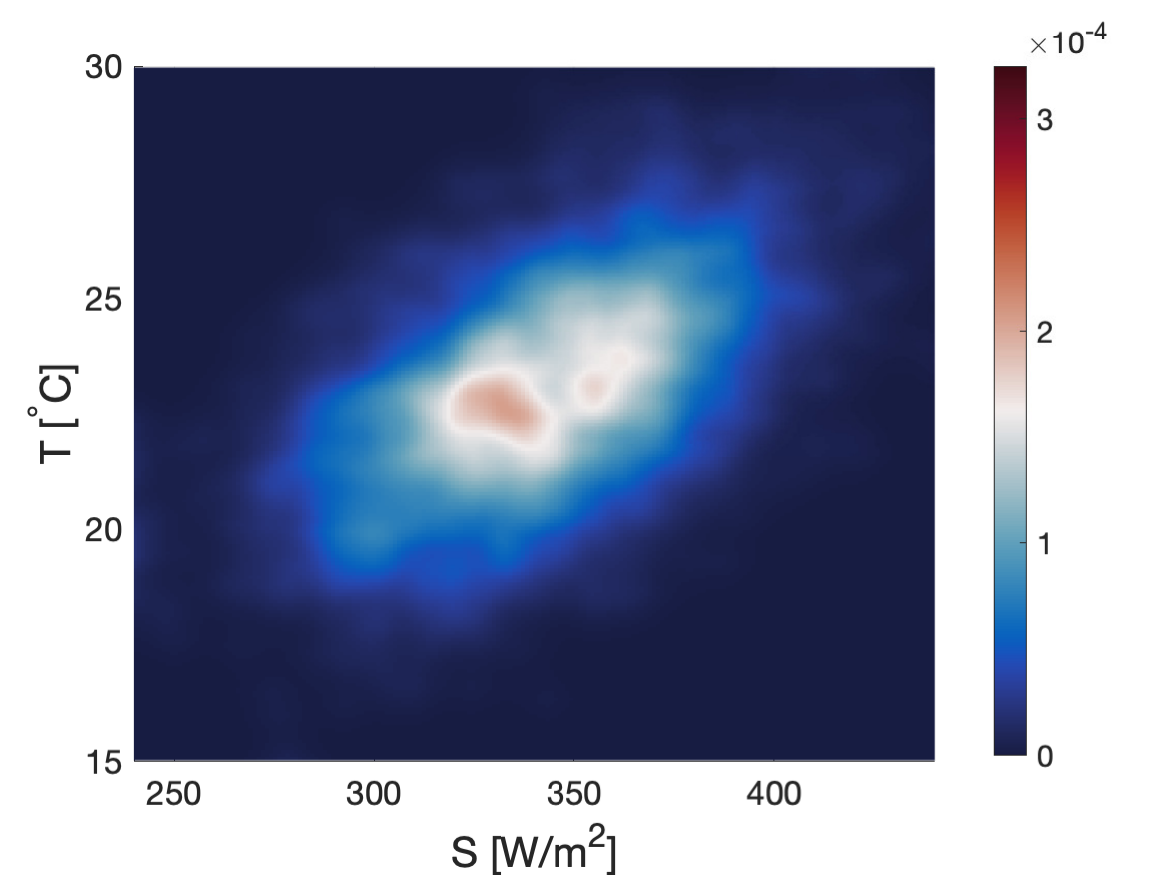}
\noindent\includegraphics[width=0.48\textwidth]{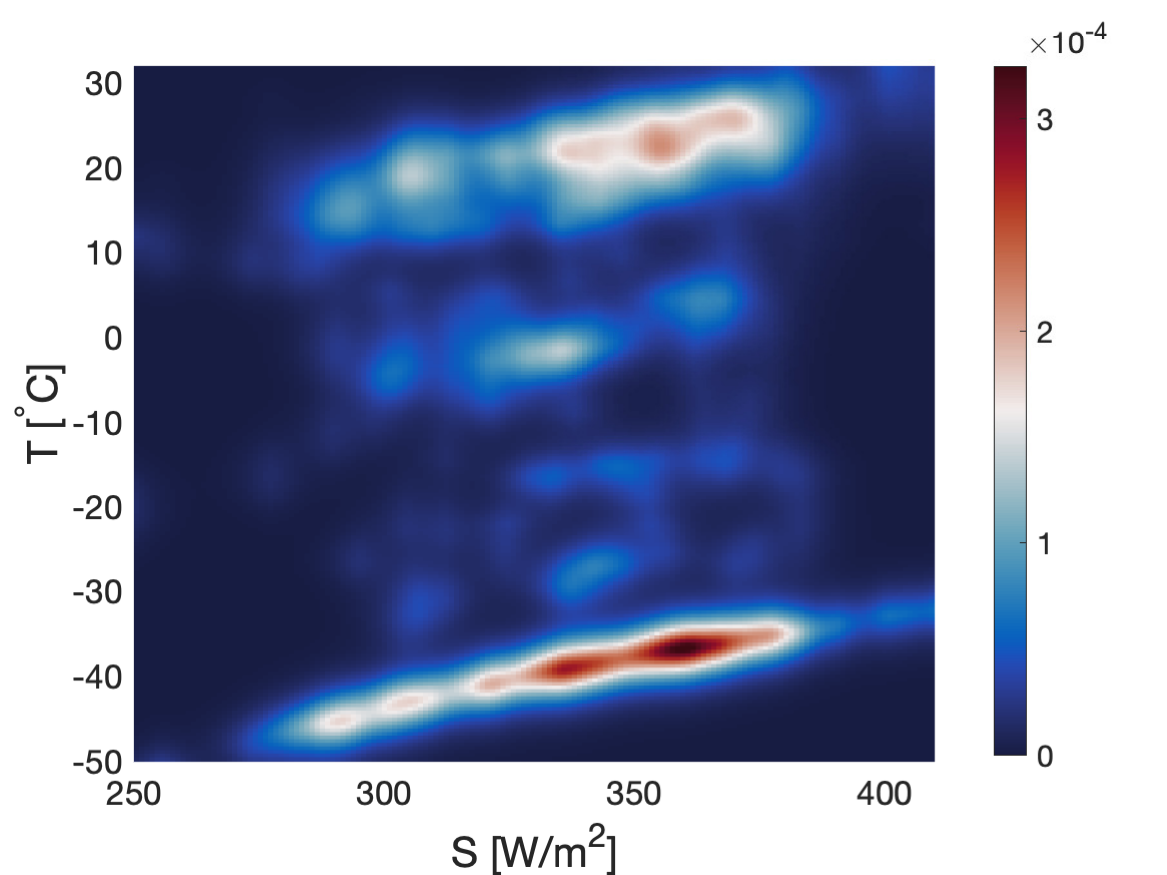}
\caption{\label{figS1} Normalised 2-dimensional histograms obtained using Method I in the main text. Random fluctuations with standard deviation  $\sigma$ are added to the incoming solar radiation at regular temporal intervals $\Delta t_1$: $\sigma = 0.025$, $\Delta t_1 = 1$~yr ({\it up left}) ;  $\sigma = 0.1$, $\Delta t_1 = 1$~yr  ({\it up right}); 
$\sigma = 0.1$, $\Delta t_1 = 10$~yr ({\it bottom}). The diverging color map goes from low (blue) to high density (red).}
\end{figure*}

\begin{figure*}
\noindent\includegraphics[width=0.49\textwidth]{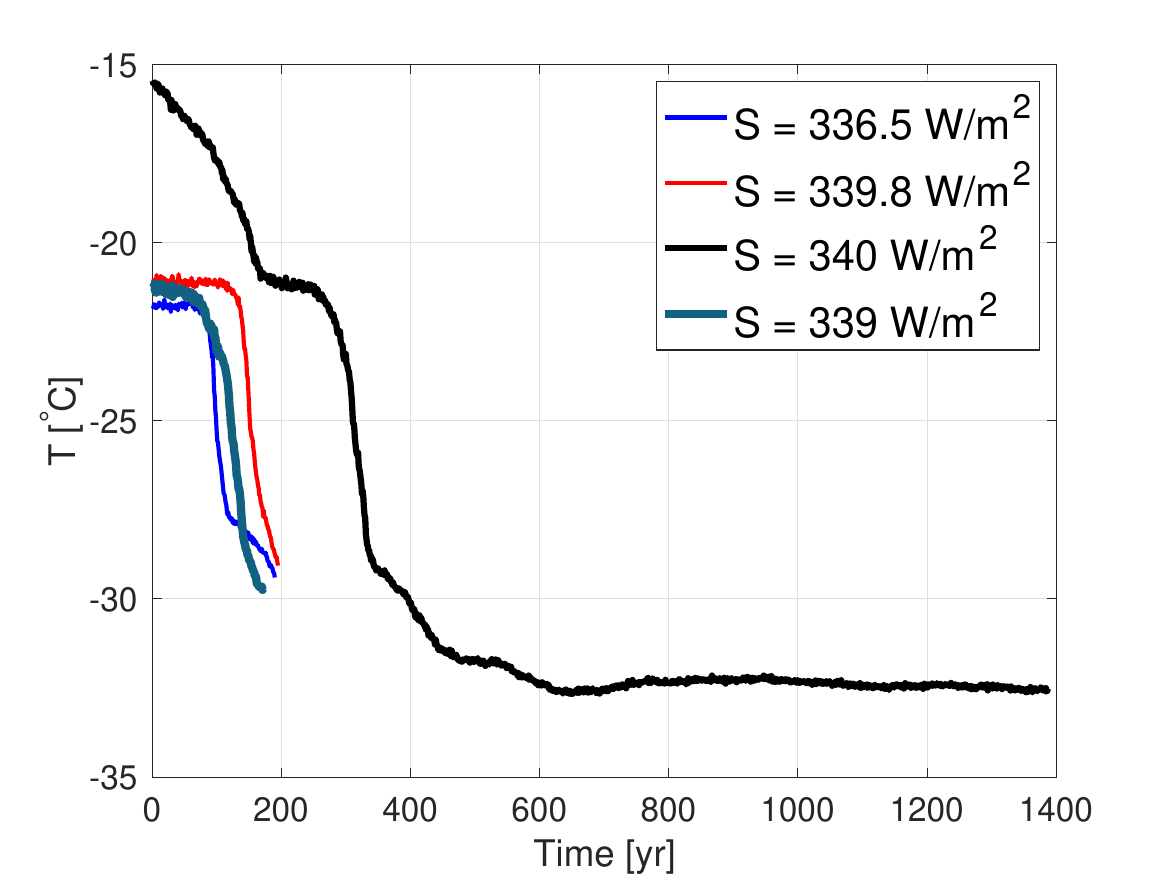}
\noindent\includegraphics[width=0.49\textwidth]{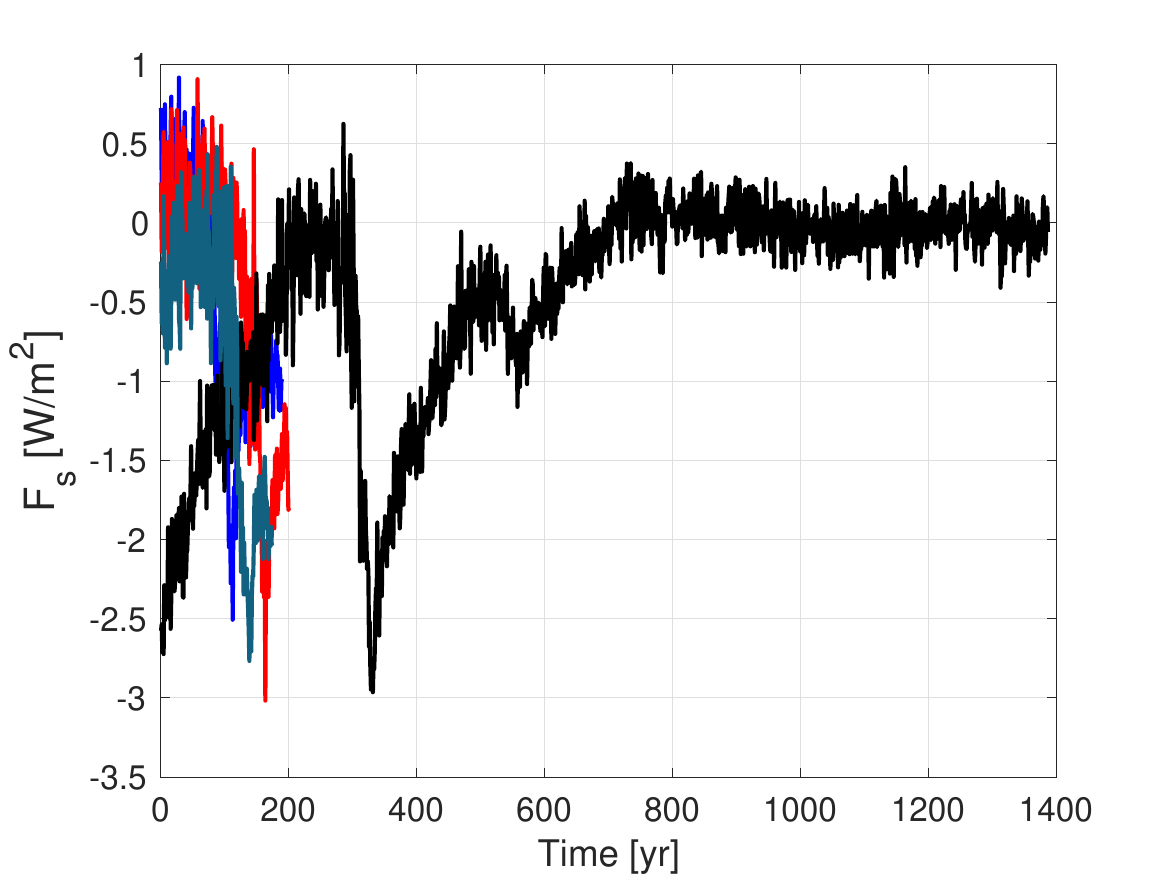}
\caption{\label{figS2} Temporal evolution of global mean surface temperature $T$ ({\it left}) and surface energy imbalance $F_{\rm s}$ ({\it right}) starting from initial conditions in the high density region near $T= -20~^\circ$C in Fig.~3 in the main article under constant incoming solar radiation $S$ (see legend). This region does not represent a new attractor since the system remains near $T= -20~^\circ$C with $F_{\rm s} \sim 0$ for only 100 years in each case to finally relax to the waterbelt climate.}
\end{figure*}

\begin{figure*}
\noindent\includegraphics[width=0.6\textwidth]{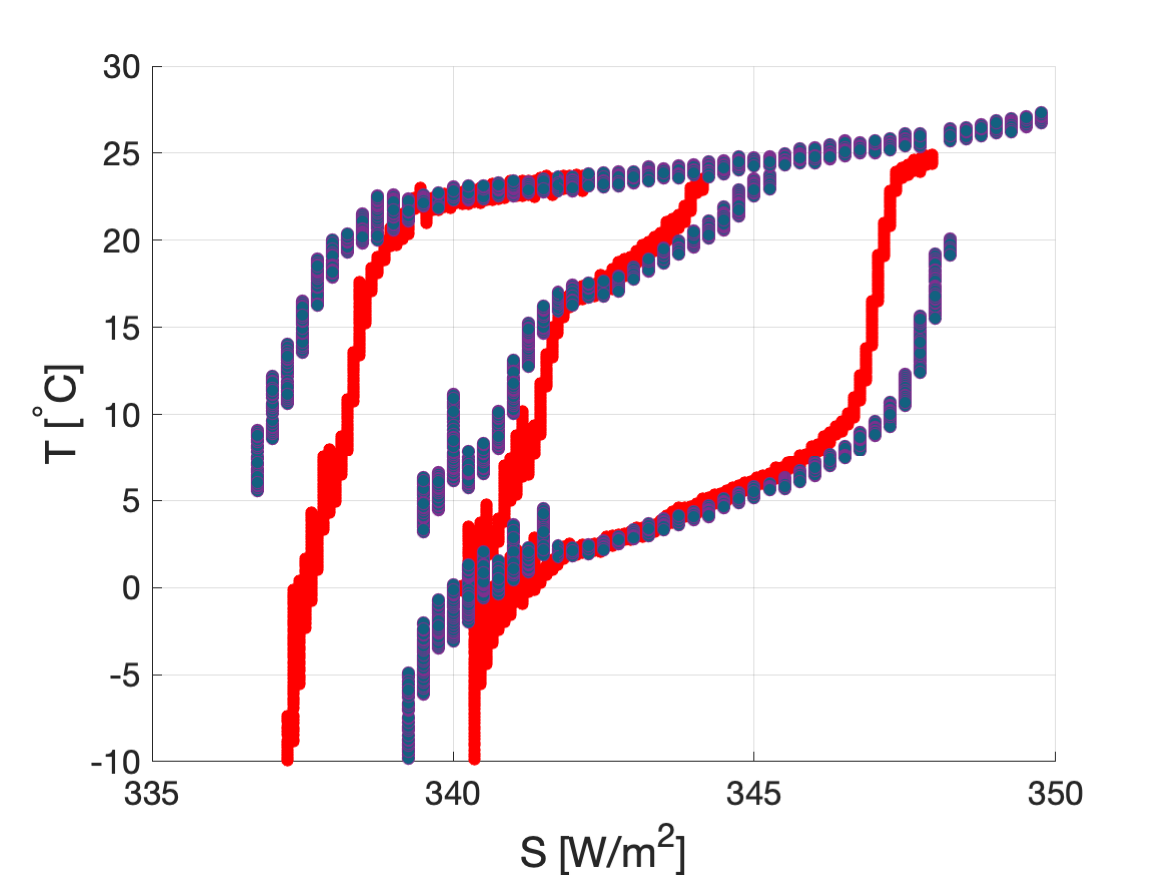}
\caption{\label{figS3} Portion of the bifurcation diagram obtained with Method II in the main text using $\Delta t_2 = 100$~yr and $\Delta S = 0.25$~W m$^{-2}$ ({\it blue}); $\Delta S = 0.1$~W m$^{-2}$ ({\it red}). The position of B-tipping is better defined with lower $\Delta S$.}
\end{figure*}

\begin{figure*}
\noindent\includegraphics[width=0.6\textwidth]{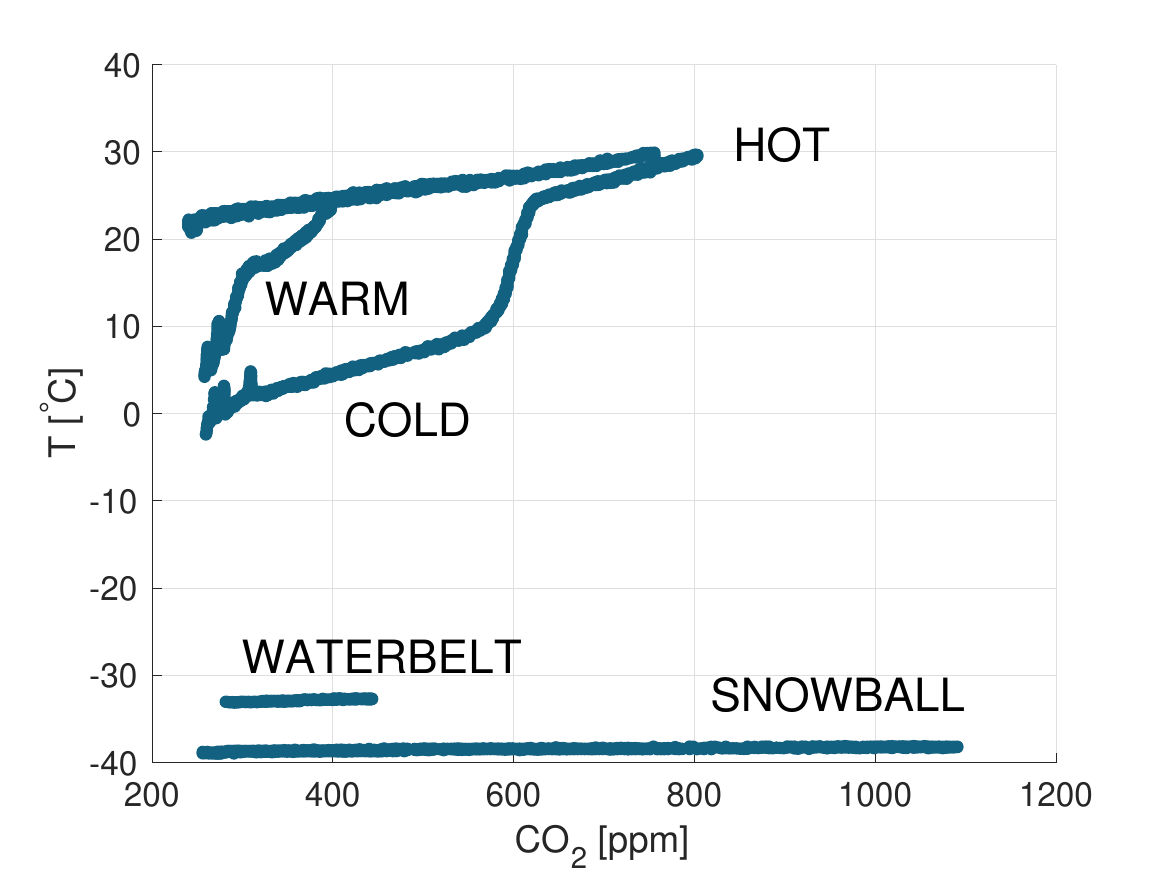}
\caption{\label{figS4} Bifurcation diagram obtained with Method II in the main text by varying the atmospheric CO$_2$ content at fixed incoming solar radiation $S_0 = 342$~W m$^{-2}$ using $\Delta t_2 = 10$~yr and $\Delta CO_2 = 0.5$~ppm. }
\end{figure*}

We provide additional figures obtained by using different parameters in the methods described in the main article (see Figures~\ref{figS1} and \ref{figS3}). We also show in Figure~\ref{figS2} that the high density region around $T \sim -20~^\circ$C in Fig.~3 does not correspond to a new attractor but to a transient feature. Finally, the bifurcation diagram obtained in terms of the atmospheric CO$_2$ content at fixed incoming solar radiation is shown in Figure~\ref{figS4}. 

%%%%%%%%%%%